\documentclass[twocolumn, switch]{article} %

\usepackage{preprint}

\usepackage{amsmath, amsthm, amssymb, amsfonts}

\usepackage[authoryear]{natbib}
\usepackage[utf8]{inputenc}	%
\usepackage[T1]{fontenc}	%
\usepackage{xcolor}		%
\usepackage[colorlinks = true,
            linkcolor = purple,
            urlcolor  = blue,
            citecolor = cyan,
            anchorcolor = black]{hyperref}	%
\usepackage{booktabs} 		%
\usepackage{nicefrac}		%
\usepackage{microtype}		%
\usepackage{lineno}		%
\usepackage{float}			%

\usepackage{lipsum}		%

\usepackage{newfloat}
\DeclareFloatingEnvironment[name={Supplementary Figure}]{suppfigure}
\usepackage{sidecap}
\sidecaptionvpos{figure}{c}

\usepackage{titlesec}
\titlespacing\section{0pt}{12pt plus 3pt minus 3pt}{1pt plus 1pt minus 1pt}
\titlespacing\subsection{0pt}{10pt plus 3pt minus 3pt}{1pt plus 1pt minus 1pt}
\titlespacing\subsubsection{0pt}{8pt plus 3pt minus 3pt}{1pt plus 1pt minus 1pt}

\usepackage{tikz,xcolor,hyperref}
\usepackage{longtable}

\definecolor{lime}{HTML}{A6CE39}
\DeclareRobustCommand{\orcidicon}{
	\begin{tikzpicture}
	\draw[lime, fill=lime] (0,0) 
	circle [radius=0.16] 
	node[white] {{\fontfamily{qag}\selectfont \tiny ID}};
	\draw[white, fill=white] (-0.0625,0.095) 
	circle [radius=0.007];
	\end{tikzpicture}
	\hspace{-2mm}
}
\foreach \x in {A, ..., Z}{\expandafter\xdef\csname orcid\x\endcsname{\noexpand\href{https://orcid.org/\csname orcidauthor\x\endcsname}
			{\noexpand\orcidicon}}
}

\title{SeisBench - A Toolbox for Machine Learning in Seismology}

\usepackage{authblk}

\author[1, *, \#]{Jack Woollam}
\author[2, 3, \#]{Jannes Münchmeyer}
\author[2, 7]{Frederik Tilmann}
\author[1]{Andreas Rietbrock}
\author[6]{Dietrich Lange}
\author[2]{Thomas Bornstein}
\author[4]{Tobias Diehl}
\author[5]{Carlo Giunchi}
\author[4]{Florian Haslinger}
\author[8, 9]{Dario Jozinović}
\author[8]{Alberto Michelini}
\author[2]{Joachim Saul}
\author[2]{Hugo Soto}

\affil[1]{\footnotesize{Geophysical Institute (GPI), Karlsruhe Institute of Technology, Karlsruhe, Germany}}
\affil[2]{\footnotesize{Deutsches GeoForschungsZentrum GFZ, Potsdam, Germany}}
\affil[3]{\footnotesize{Institut für Informatik, Humboldt-Universität zu Berlin, Berlin, Germany}}
\affil[4]{\footnotesize{Swiss Seismological Service, ETH Zurich, Zurich, Switzerland}}
\affil[5]{\footnotesize{Istituto Nazionale di Geofisica e Vulcanologia, Sezione di Pisa, Pisa, Italy}}
\affil[6]{\footnotesize{GEOMAR Helmholtz Centre for Ocean Research Kiel, Kiel, Germany}}
\affil[7]{\footnotesize{Institut für geologische Wissenschaften, Freie Universität Berlin, Berlin, Germany}}
\affil[8]{\footnotesize{Istituto Nazionale di Geofisica e Vulcanologia, Roma, Italy}}
\affil[9]{\footnotesize{Unversit\`a degli Studi Roma Tre, Largo San Leonardo Murialdo 1, Rome, Italy}}
\affil[*]{\footnotesize{corresponding author}}
\affil[$\#$]{\footnotesize{equal contribution}}

\begin{document}

\twocolumn[ %
  \begin{@twocolumnfalse} %
  
\maketitle

\begin{abstract}
Machine Learning (ML) methods have seen widespread adoption in seismology in recent years. 
The ability of these techniques to efficiently infer the statistical properties of large datasets often provides significant improvements over traditional techniques when the number of data are large(>>millions of examples).
With the entire spectrum of seismological tasks, e.g., seismic picking and detection, magnitude and source property estimation, ground motion prediction, hypocentre determination; among others, now incorporating ML approaches, numerous models are emerging as these techniques are further adopted within seismology. 
To evaluate these algorithms, quality controlled benchmark datasets that contain representative class distributions are vital. 
In addition to this, models require implementation through a common framework to facilitate comparison. 
Accessing these various benchmark datasets for training and implementing the standardization of models is currently a time-consuming process, hindering further advancement of ML techniques within seismology. These development bottlenecks also affect 'practitioners' seeking to deploy the latest models on seismic data, without having to necessarily learn entirely new ML frameworks to perform this task.
We present SeisBench as a software package to tackle these issues. SeisBench is an open-source framework for deploying ML in seismology. 
SeisBench standardises access to both models and datasets, whilst also providing a range of common processing and data augmentation operations through the API. 
Through SeisBench, users can access several seismological ML models and benchmark datasets available in the literature via a single interface. 
SeisBench is built to be extensible, with community involvement encouraged to expand the package. 
Having such frameworks available for accessing leading ML models forms an essential tool for seismologists seeking to iterate and apply the next generation of ML techniques to seismic data.
\end{abstract}
\vspace{0.35cm}

  \end{@twocolumnfalse} %
] %

\section{Introduction}

Seismology has always been a ‘data-rich’ field. 
With the continued advances in computational power and decreasing cost of high quality seismometers, the seismic wavefield is now recorded with increasing resolution and fidelity. 
Such advances are not just exclusive to seismology; within science in general, larger, more detailed datasets are being compiled. 
Machine Learning (ML) has risen to prominence as a set of techniques to best exploit the information contained in such extensive datasets.
Often termed ‘data-driven’ methods, ML tools probabilistically model the statistical properties of a given dataset to perform inference for a given task.
As datasets get larger, and the inference step becomes a more tractable problem, these techniques are now achieving state-of-the-art performance across the entire spectrum of scientific fields. 
In many areas performance is outpacing the human analyst.

Although some pioneering works harnessed neural networks for seismological applications, for many years \citep[e.g][]{wang1997identification,valentine2012data} such techniques did not find wider usage in seismology until approximately three years ago.  
The possibility to assemble large datasets, massive parallelisation on commodity hardware through GPU computing, algorithmic improvements and, importantly, the availability of software frameworks such as PyTorch \citep{paszke2017automatic} and Tensorflow \citep{abadi2016tensorflow} has driven a wave of applications of ML techniques to classical seismological problems, including earthquake phase identification \citep{zhu2019phasenet, ross2018generalized, woollam2019convolutional, zhou2019hybrid, wang2019deep}, earthquake detection \citep{mousavi2019cred, mousavi2020earthquake, zhou2019hybrid, perol2018convolutional, dokht2019seismic, pardo2019seismic}, magnitude estimation \citep{lomax2019investigation, mousavi2020machine, van2020automated, munchmeyer2021teamlm}, and earthquake early warning \citep{kong2016myshake, li2018machine, munchmeyer2021team}, amongst others. 

From these initial works, a natural question arises: which ML techniques perform best for each task? 
Answering this question is not trivial, as each study uses different data, different ML frameworks for algorithm development, and different assessment metrics.
Benchmarking and comparison studies are, therefore, inherently difficult.
The varying data used during training is a particular problem, as the variable nature of earthquake source, propagation medium and site conditions mean that the performance of a model trained on one region or environment might not be directly compared to a model trained in a different region.
To enable fair comparisons of models and model architectures over a range of possible environments, benchmark datasets are essential.  

Labelled benchmark datasets have been vital to rapid-progress in various classic ML application domains, most prominently 
computer-vision (MNIST, \citeauthor{deng2012mnist}, \citeyear{deng2012mnist}; ImageNet \citeauthor{deng2009imagenet}, \citeyear{deng2009imagenet}), and natural language processing (Sentiment140, \citeauthor{sentiment140}, \citeyear{sentiment140}), as they allow for easy assessment of which ML algorithms perform best. 
Creating such quality-controlled datasets takes, however, a significant amount of time. 
Benchmark datasets perform this step for users ensuring comparability of different studies, greatly accelerating the development and testing of novel ML algorithms. 
With ML methods only recently being widely adopted in seismology, historically, there were no benchmark datasets available for comparison works. 
This situation is now changing with the value of such datasets widely recognised. 
The seismological benchmark datasets now emerging (e.g. LenDB, \citeauthor{magrini2020local}, \citeyear{magrini2020local}; INSTANCE, \citeauthor{michelini2021instance}, \citeyear{michelini2021instance}; NEIC, \citeauthor{yeck2021leveraging}, \citeyear{yeck2021leveraging}; STEAD, \citeauthor{mousavi2019stanford}, \citeyear{mousavi2019stanford}) already cover a wide-range of potential seismic environments (e.g. global, regional, local), essential factors for training robust algorithms.

However, the availability of new benchmark datasets does not completely solve the comparison problem.
Remaining issues include the differing data formats employed by different benchmarks, and the specific framework libraries ML researchers use to implement their models e.g. PyTorch, Tensorflow, Keras \citep{chollet2015keras}, Sklearn \citep{pedregosa2011scikit}, add complexity to any comparison work. 
Any benchmarking must, therefore, check that operations applied within each library are directly comparable, with no discrepancies in implementation.

The easy availability of both benchmark datasets, and standardised access to the latest models, are crucial ingredients for advancing the state-of-the-art. 
As this problem is common to any application based on ML \citep{kiraly2021designing}, tools have been developed in other fields to provide researchers with easy access to models and benchmark datasets (e.g. FLAIR, \citeauthor{akbik2019flair}, \citeyear{akbik2019flair}, natural language processing toolbox). 
These continue to be widely used, evidence of their ability to aid development. 

To date, we are unaware of the availability of such software in seismology. 
The outlined bottlenecks affect a wide range of potential users of ML. 
For the 'practitioner', who wishes to apply ML models to their seismic data, they are currently facing significant hurdles, as they would have to learn specific frameworks to integrate the latest ML algorithms into their workflows. 
For the 'expert' interested in developing novel techniques, they currently have to integrate various models, testing over varying datasets, which may be in differing formats. 
Without any frameworks or toolboxes to help with these problems, researchers must construct such comparison pipelines from scratch. 
This is a significant undertaking.
These factors are currently hindering more widespread ML adoption in seismology and are limiting progress in the development of the next generation of ML methods. 
Tackling these problems is key if the seismological community is to accelerate the development of ML techniques for seismic tasks and promote further adoption of ML within the field. We have built the SeisBench open-source software package to address these issues.

\section{The SeisBench ML framework}

\begin{figure*}
  \includegraphics[width=\textwidth]{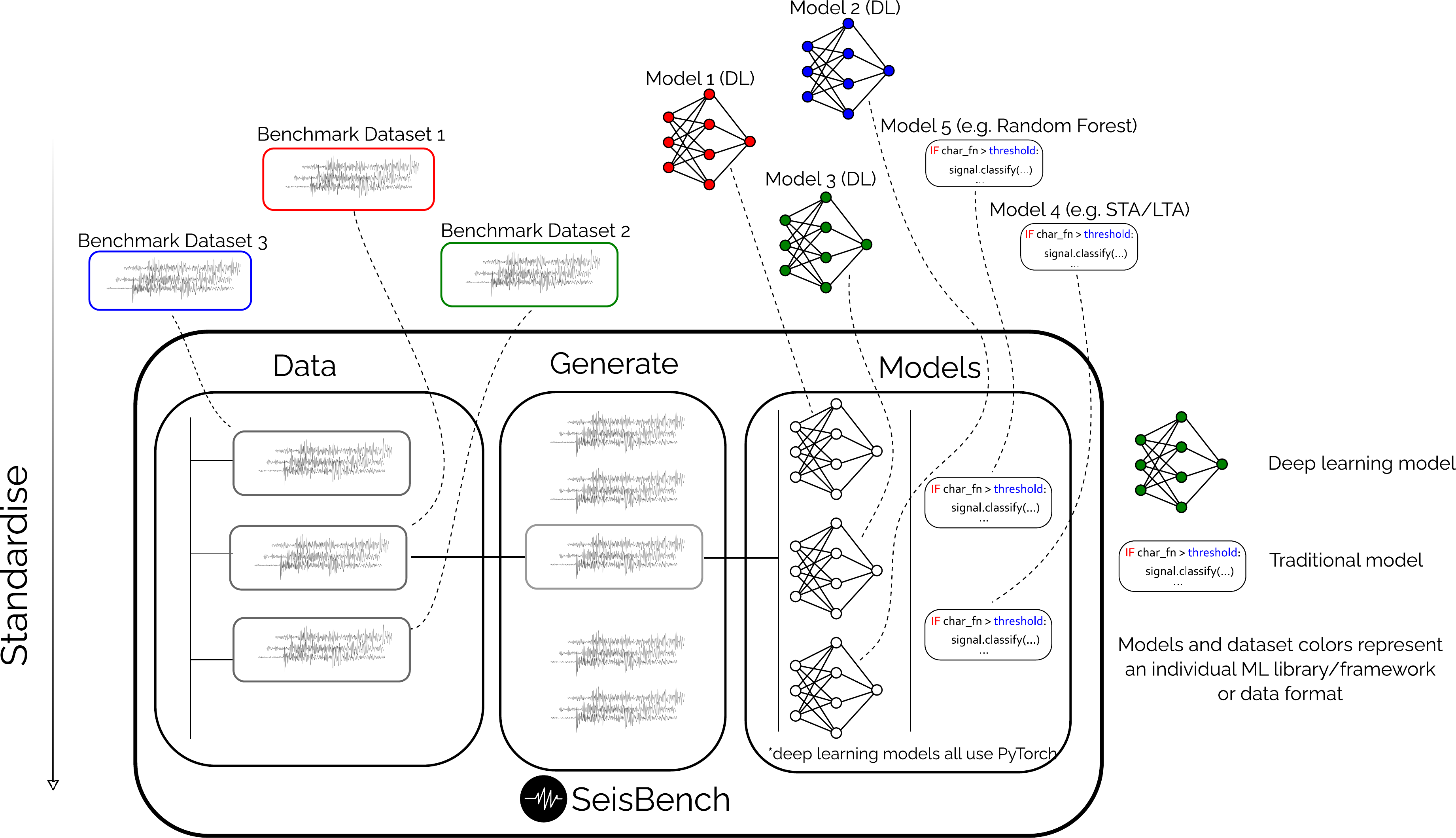}
  \caption{Schematic diagram to show the motivation behind SeisBench. SeisBench acts as a unifying framework for developing models and applying them to seismic data. The differing packages used for model development, and the differing benchmark dataset formats are represented by varying colours.}
\label{fig_sb_schematic}
\end{figure*}

SeisBench provides a unified point-of-access for ML development and application within the seismological community. 
Built in Python, it integrates both state-of-the-art models and datasets in a single framework. 
Figure \ref{fig_sb_schematic} visually highlights this concept, introducing the core components of SeisBench.
The range of datasets included in the initial release include currently published seismological benchmark datasets from the literature, directly integrated into the package. 

SeisBench also provides access to additional custom benchmark datasets which are made newly available in the initial release of the software. 
As all datasets within SeisBench adhere to a common format, users can compare their algorithms across a range of seismic environments, from detecting global signals to local settings. 
Models are accessed through a unified interface – enabling easy comparison of differing approaches. 
Whilst the model interface is designed to integrate various deep learning models, the types of models that can be built and compared in SeisBench are not just limited to deep learning algorithms; traditional methods can also be directly deployed and integrated into comparison workflows. 
Finally, typical data augmentation and pre-processing steps are provided through an augmentation API. 
With seismologists, and general ML practitioners, often re-implementing the same operations for data pre-processing and augmentation, inclusion of many of the standard processes and augmentations in SeisBench will further facilitate faster model development. 

SeisBench is designed to be generally applicable to the entire spectrum of general seismological tasks, such as source parameter estimation, magnitude estimation, ground motion prediction. 
Whilst the currently included models relate specifically to picking and event detection, SeisBench is suitable for many other seismological tasks based on waveform analysis. 
The extensible nature of the API means that any parameter from a datasets metadata can be used as a label (target variable), enabling the construction of any supervised classification pipeline.

\subsection{Data - Standardising access to Benchmark datasets}

\subsubsection{A standardised format for seismic waveforms and metadata information}

\begin{figure*}
  \centering
  \includegraphics[width=0.7\textwidth]{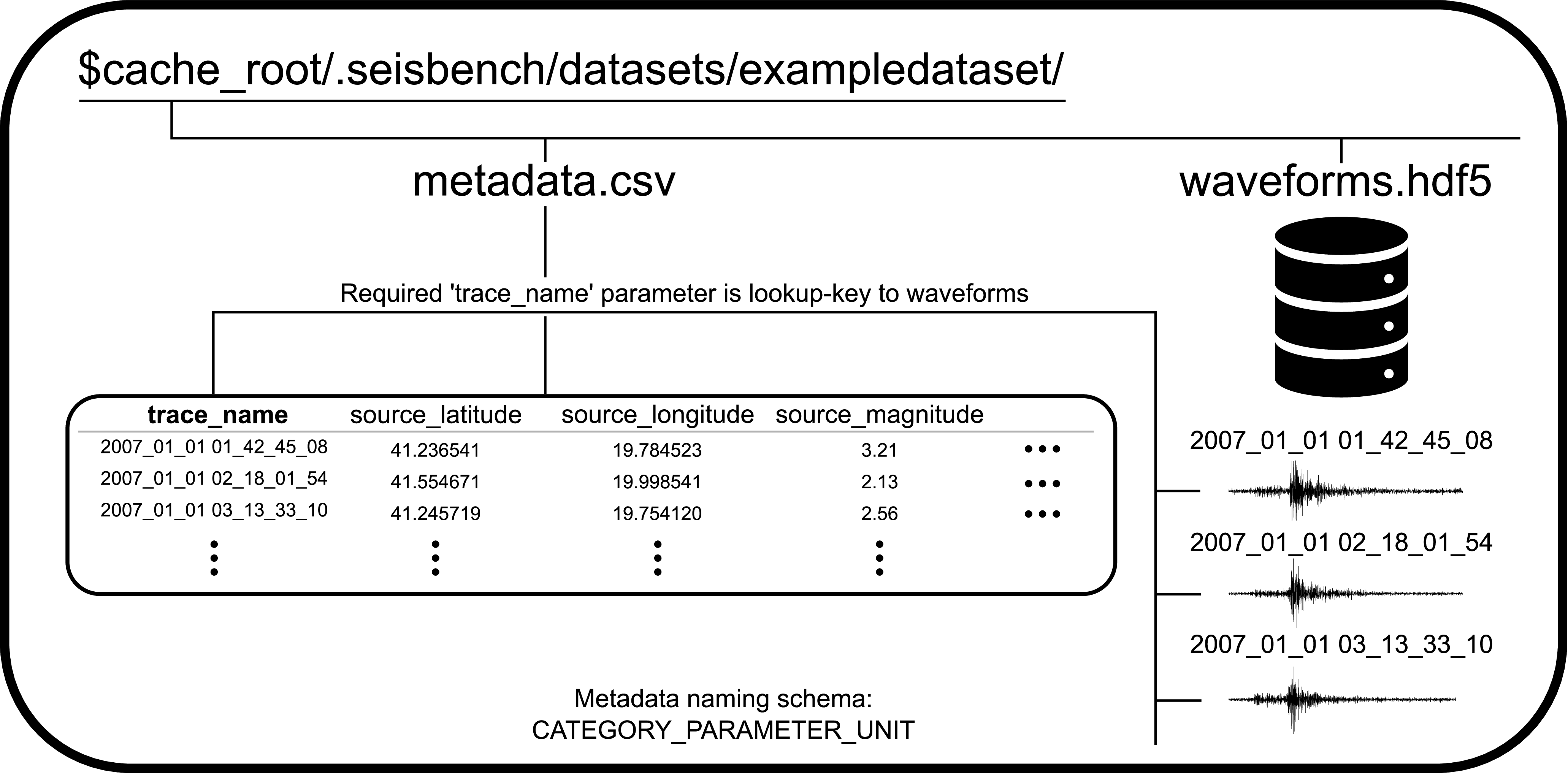}
  \caption{Example of data structure for SeisBench. Waveforms are stored in a HDF5 file, indexed by trace name. The metadata for each waveform example is stored in a table format as a .csv file. The trace name is required as a column, as this is then used as the lookup key to the raw data. This schematic diagram displays the overall concept, with the implementation slightly more complex to optimise performance. For more information see the technical documentation (\url{https://seisbench.readthedocs.io/en/latest/)}.}
  \label{fig_data_format}
\end{figure*}

The SeisBench \textit{data} module contains functionality to read and process seismological datasets which have been converted into the SeisBench standardised format. Using a standardised framework enables the construction or conversion of varying benchmark seismological datasets. 
The dataset format follows a typical approach encountered within the ML community (Figure \ref{fig_data_format}), where the waveforms  (training examples) are included in a single file. 
We use Hierarchical Data Format 5 (HDF5) to store the raw waveforms \citep{folk2011overview}. Each multi-component waveform example is indexed by a lookup key. 
For all datasets, the required parameter \textit{‘trace\textunderscore name’} is used as the lookup key. 
The labels/metadata associated with each training example are then stored in a simple table-structure (.csv). 
To ensure compatibility across datasets, metadata parameter names should follow a common naming schema ‘CATEGORY\textunderscore PARAMETER\textunderscore UNIT’ where: category defines the object which the parameter describes (i.e., path, source, station, trace); parameter describes the provided information e.g. latitude or longitude; and unit provides the unit of measurement e.g., m, cm, s, counts, samples\footnote{An example of some of the more typically encountered metadata parameters for SeisBench datasets, and how they would be named in the SeisBench format, can be found in the Data Appendix (Table \ref{param_description_table}).}.

Where several entries are required, such as trace start time and station name and location, such a data structure leaves the freedom to include additional specialised metadata only available for selected datasets. 
The metadata information is read into memory with the popular, high-level data-analysis library Pandas \citep{reback2020pandas}. 
With such a format, users can easily create their own custom pipelines to query and extract metadata information associated with the waveforms. 
Providing a common framework for data storage is key to any proposed benchmarking works. 
Imposing restrictions on both the format and naming schema ensures that any newly defined parameters are still standardised across datasets. 
This greatly aids extensibility and comparability across datasets. 
Data throughput can be a major factor in the efficiency of training and application of ML models. SeisBench therefore introduces additional performance optimizations to the data structure that enhance IO read/write speed.  

Once a dataset has been converted to the SeisBench format, it is integrated into the SeisBench API by extending the base dataset interface, providing a unique class for the dataset. 
Ordering the datasets into a class-based hierachy naturally reflects the dataset format. 
Common operations such as filtering metadata and obtaining waveforms are all available via the base dataset interface. 
Further individual properties of each dataset can then be encapsulated in the dataset class.  
Tools are available to help scientists to convert their own datasets into benchmark datasets and contribute them to the SeisBench repository, if desired. 

\subsubsection{Providing a common endpoint for benchmark datasets}
\begin{figure*}
  \centering
  \includegraphics[width=\textwidth]{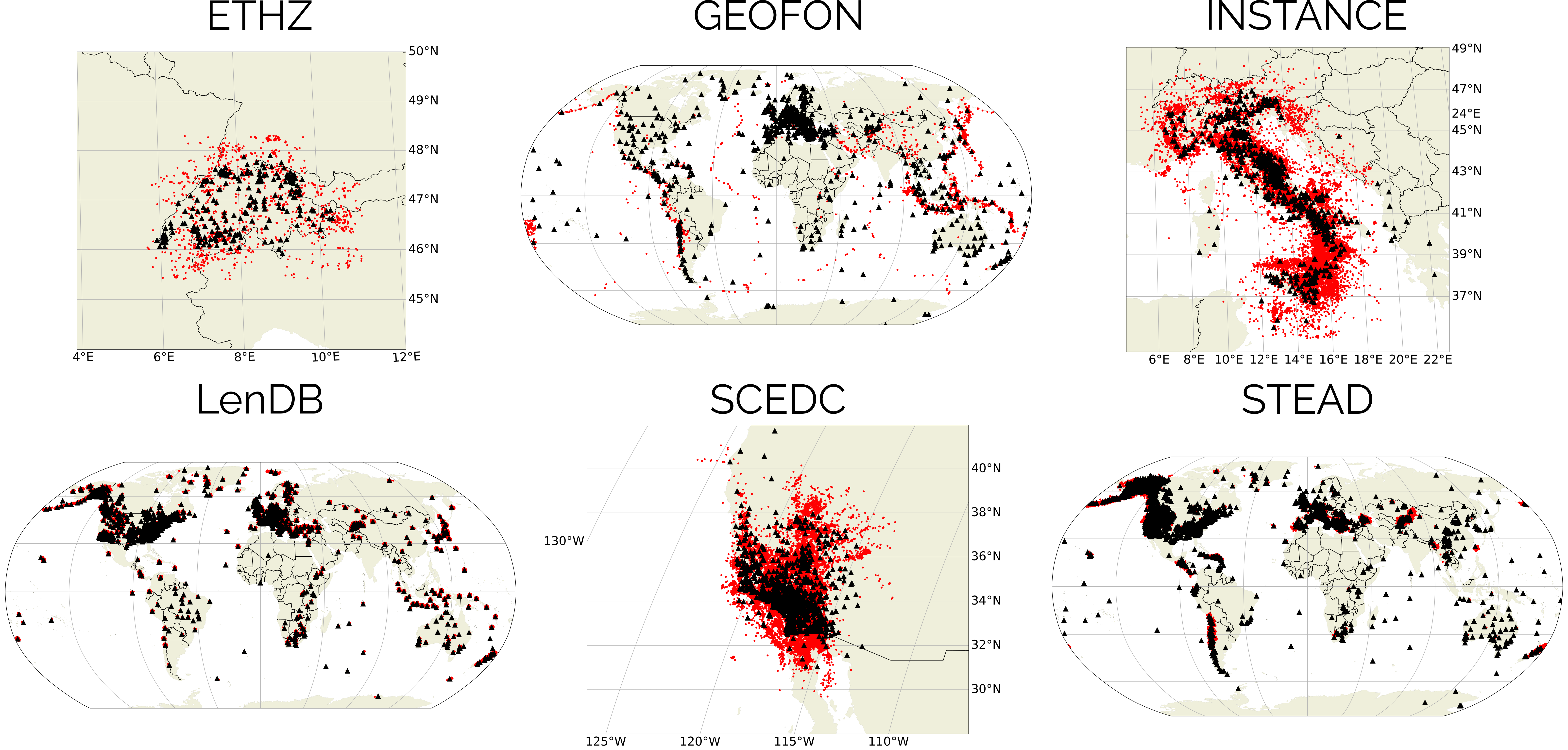}
  \caption{Benchmark datasets integrated into SeisBench with the initial release of the software; seismic sources are circles, stations are triangle markers. Not shown are some additional datasets which are included in the SeisBench initial release dataset collection, but are either missing missing source information (NEIC, GPD, JGRPick, JGRFM, Meier2019JGR), or have minimal number of events for plotting (the local Iquique dataset).}
\label{fig_datasets}
\end{figure*}

\begin{figure*}
  \includegraphics[width=\textwidth]{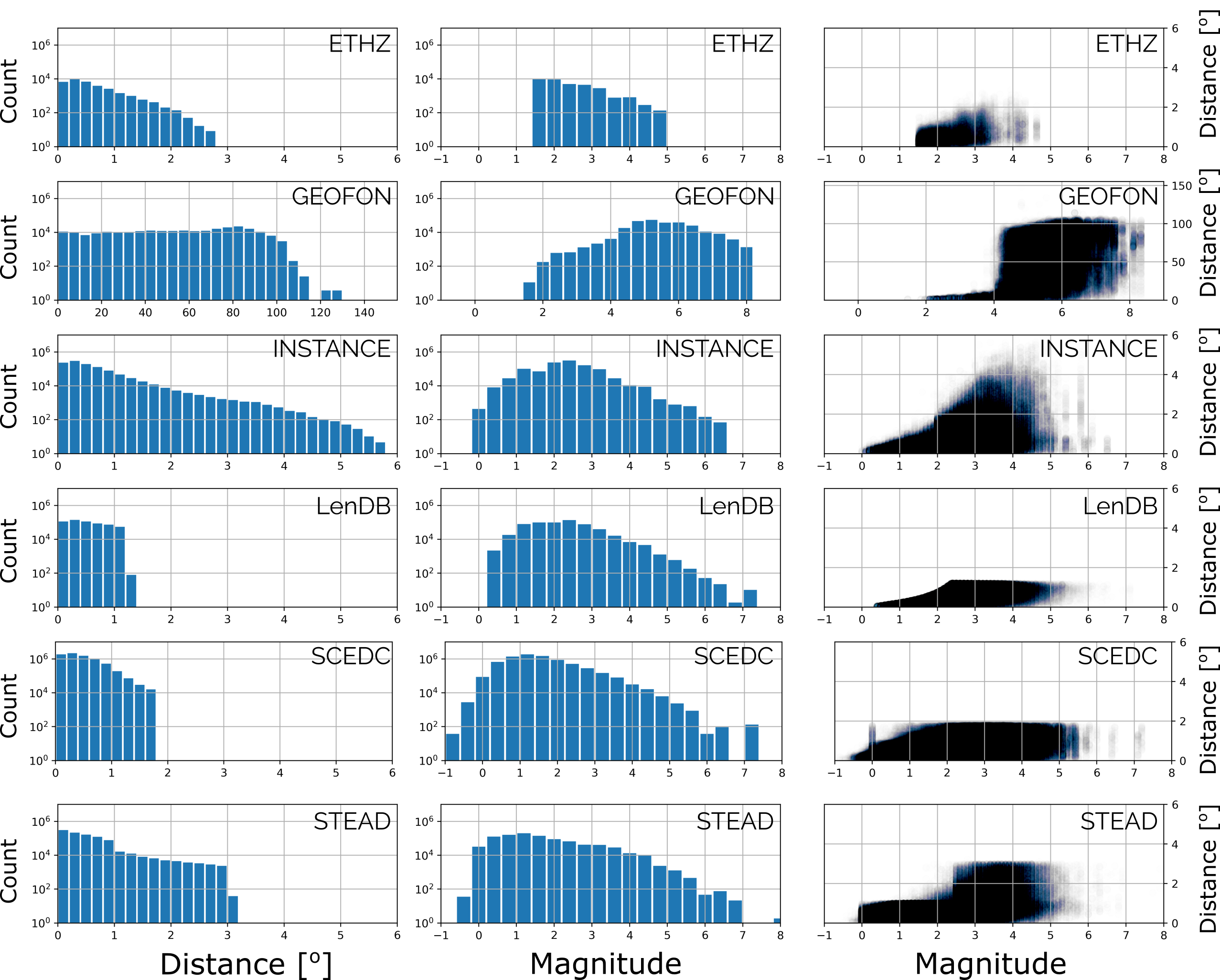}
  \caption{Logarithmic histograms of epicentral distance  and magnitude distributions for the datasets with source and station information. For the two-dimensional scatterplot in the last column, all points are plotted with transparency to highlight the overall distribution. The Iquique, NEIC, GPD, JGRPick, JGRFM, Meier2019JGR datasets are not shown because they are lacking either, magnitude, or source and station location information.}
\label{fig_datasets_info}
\end{figure*}

We have converted a range of seismological benchmark datasets (Table \ref{dataset_table}; Figure \ref{fig_datasets}) into the SeisBench data format. These datasets contain various types of seismic arrivals from local to global scales (Figure \ref{fig_datasets_info}).
All the datasets were either compiled from publicly available seismic data and metadata, or were directly converted from a published benchmark dataset from the literature. 
SeisBench thus provides easy access to data and model interfaces. All users have to do upon installation of the package is to instantiate their preferred data/model object; the data will then be downloaded and cached for repeat use.
Within each benchmark dataset, training, validation, and testing splits are pre-defined in order to reduce variability of benchmark comparisons resulting from randomness or different choices for dataset splitting approaches. Of course, it remains possible to define custom splits for specialized applications.  
Here, we summarise the benchmark datasets integrated into the first release of SeisBench. The benchmark datasets can be separated into two groups, datasets that are missing some common metadata such as station location information, and those that contain all typical metadata information such as the station location and source parameters. Table \ref{dataset_table}, and Figure \ref{fig_datasets} and \ref{fig_datasets_info} generally only show those datasets of the first group, where all the common metadata are present.
The following dataset descriptions provide further information on the included metadata.

\begin{table*}
 \centering
 \caption{Overview of the datasets. The noise column indicates the number of dedicated noise traces. Note that it is still possible to extract noise examples from datasets without dedicated noise traces by selecting windows before the first arrival. For distances, the datasets cover local(L, 0 $\leq$ $\Delta$ \textless 150~km), regional (R, 150 $\leq$ $\Delta$ \textless 600~km), and teleseismic (T, $\Delta$ > 600~km) records. The datasets with variable trace length contain considerably more than 60s of data for most examples. $f_s$ denotes the sampling rate. When this parameter varies within a dataset,  the range of sampling rates is listed. The GPD, JGRPick, JGRFM, Meier2019JGR datasets are omitted from this table because these datasets do not contain source property information. NEIC is included because it is used for the benchmark comparison in \protect\citet{munchmeyer2021benchmark}.}
\setlength{\tabcolsep}{3pt}
\footnotesize
\begin{tabular}{c|c|c|c|c|c|c|c|c|c|}
& Traces & Events & P picks & S picks & Noise & Region & Dist & Tr. length & $f_s$ [Hz]\\
\hline
ETHZ & 36,743 & 2,231 & 35,227 & 18,960 & 0 & Switzerland & L/R & variable & 100 - 500\\
INSTANCE & 1,291,537 & 54,008 & 1,159,249 & 713,883 & 132,288 & Italy & L/R & 120~s & 100\\
Iquique & 13,400 & 409 & 13,327 & 11,361 & 0 & N. Chile & L & variable & 100\\
LenDB & 1,244,942 & 303,902 & 629,095 & 0 & 615,847 & various & L & 27~s & 20\\
SCEDC & 8,111,060 & 378,528 & 7,571,970 & 4,364,155 & 0 & S. California & L & variable & 40 - 100\\
STEAD & 1,265,657 & 441,705 & 1,030,231 & 1,030,231 & 235,426 & various & L/R & 60~s & 100\\
\hline
GEOFON & 275,274 & 2,270 & 284,240 & 2,847 & 0 & global & R/T & variable & 20 - 200\\
NEIC & 1,354,789 & 137,424 & 1,025,000 & 329,789 & 0 & global & R/T & 60~s & 40\\
\end{tabular}
\label{dataset_table}
\end{table*}

\noindent \\ETHZ\\
\noindent The ETHZ benchmark dataset is a manually compiled dataset for SeisBench. 
It contains local to regionally recorded seismicity throughout Switzerland and neighbouring border regions. 
The data are recorded on the publicly available networks: 
8D \citep{sed8d}, C4 \citep{sedc4}, CH \citep{sedch}, S \citep{seds}, XT \citep{sedxt_2014}, operated by the Swiss Seismological Service (SED) at ETH Zurich. 
To construct this dataset, we obtained both the waveform recordings and the corresponding metadata information via SED's FDSN web service (\url{http://www.seismo.ethz.ch/de/research-and-teaching/products-software/fdsn-web-services/}). 
Any detected seismic event from this network has had the phases manually labelled, including the discrimination of first, and later phases (e.g. Pn vs. Pg). 
In addition to the typical phase identification, the magnitude and polarity information is also available. 
In total, there are 57 metadata variables available for this dataset. 
We select all M > 1.5 events from the period of 2013 - 2020 for integration. 
In total there are 2,231 events containing 36,743 waveform examples. 
The traces are all in raw counts.

We split training examples for this dataset into training, development, and testing example splits by setting all events before August 1st 2019 as training examples (61.6\%), all events between this date and the 4th September 2019 are set as the development split (9.9\%), and all the remaining events later than this date are the testing split (28.5\%).\\

\noindent GEOFON\\
\noindent The GEOFON monitoring service acquires and analysis waveforms from over 800, globally distributed seismic stations worldwide. 
The GEOFON benchmark dataset has been compiled from these recordings. 
It is a teleseismic dataset which includes 2270 events containing $\sim$275,000 waveform examples occurring between 2009 – 2013. 
Events have been picked automatically initially, with manual analysis and onset re-picking performed routinely whenever necessary to improve the location quality. 
The magnitudes range from $\sim$M 2 - 9. With the bulk of events compromising intermediate to large events (M 5-7; Figure \ref{fig_datasets_info}). 
Any regional events with smaller magnitudes are predominantly from the regions of Europe and northern Chile. 
54 metadata variables are included with this dataset, the trace units are in raw counts. 

For the GEOFON dataset, please note the varying class distributions of picked phase types for this dataset. 
For local and near-regional events S onsets have been picked and for a small fraction both Pn and Pg are included. 
For teleseismic events, almost no S onsets have been picked. 
Depth phases have been picked occasionally but not comprehensively.

For the training, validation and testing splits, we set all events occurring before 1st November 2012 as training examples (58.6\%), all events between this date and 15th March 2013 as the validation examples (10.1\%), and any remaining events past this date as the testing examples (31.3\%).\\

\noindent INSTANCE\\
\noindent The INSTANCE benchmark dataset \citep{michelini2021instance} comprises $\sim$1.3 million regional 3-component waveforms from $\sim$50,000 earthquakes M 0 – 6.5 and also includes $\sim$130,000 noise examples. 
Within SeisBench, we provide separate access to the individual partitions of this dataset. 
The noise examples and signal examples are available as their own distinct dataset; 
the seismic events are further subdivided into datasets with waveforms in counts, and with waveforms in ground motion units. 
A combined dataset containing all noise examples and waveform examples in counts is also available. 
A total of 115 metadata variables are provided. 
In addition to the standard metadata variables, this dataset includes a rich set of derived metadata, e.g. peak ground acceleration and velocity.

The training, validation, and testing sets are performed by randomly selecting 'event-wise' for this dataset. 
All waveform examples belonging to the same event are, therefore, in the same split group. 
The final proportion of waveform examples for each class are 60.3\% for training, 10\% for validation, and 29.7\% for testing respectively.\\

\noindent Iquique\\
\noindent The Iquique benchmark dataset is a benchmark dataset of locally recorded seismic arrivals throughout northern Chile originally used in training the deep learning picker in \cite{woollam2019convolutional}. 
It contains 13,400 waveform examples with 13,327 manual P-phase picks and 11,361 manual S-phase picks. 
All waveform units are in raw counts, there are 23 metadata variables associated with this dataset. 

For this dataset, the training, validation and testing splits are selected through randomly sampling the training examples, returning 60\%, 30\% and 10\% for the training, validation, and testing splits respectively.

\noindent LENDB\\
\noindent The LENDB benchmark dataset \citep{magrini2020local} is a published benchmark dataset containing local earthquakes recorded across a global set of 1487 broad-band and very broad-band seismic stations. 
It comprises $\sim$1.25 million waveforms. 
The dataset is split into 629,095 local earthquake examples and 615,847 noise examples. 
The data were processed using a bandpass filter between 0.1 - 5~Hz and the instrument response was deconvolved to convert the recordings into physical units of velocity. 
Unlike the other datasets, only automatic P-phase picks are provided for LenDB. 
In total there are 23 metadata variables for this dataset.

The training, validation, testing split is performed by selecting all examples with waveform start times before 16th January 2017 as training examples (60\%). Any examples between this date and the 16th August 2017 form the validation split (9.5\%), and the remaining examples past this date form the test split (30.5\%).\\

\noindent SCEDC\\
\noindent The Southern Californian Earthquake Data Centre (SCEDC) benchmark dataset has been constructed from publicly available waveform data \citep{scedc2013southern}. The 
waveforms and associated metadata are obtained via the Seismic Transfer Programme (STP) client \citep{stpsoftware}. 
For the obtained seismic arrivals, all events have been manually picked. 
We select all publicly available recordings of seismic events in the Southern Californian Seismic Network, over the period 2000 - 2020. 
Only local recordings of seismic events ($\sim$M -1 – 7) are included, with source to station paths spanning up to a maximum distance of $\sim$200~km. 
The dataset comprises $\sim$8 million waveform examples, which contain $\sim$7.5 million P-phases and $\sim$4.3 million S-phases. 
This dataset also contains a range of seismic instrument types including: extremely short period, short period, very broadband, broadband, intermediate band and long period instruments - both single and 3-component channels are also present. Units for the examples are raw counts. 

The split for this dataset is set randomly, with 60\%, 10\%, and 30\% of the data compromising the training, development, and testing splits respectively. 
For the magnitude metadata information, please note the increase of M = 0 in events in comparison to the overall trend \ref{fig_datasets_info} which suggests some data cleaning is still required for this dataset for the purposes of magnitude prediction.\\

\noindent STEAD\\
\noindent The STanford EArthquake Dataset (STEAD; \citeauthor{mousavi2019stanford}, \citeyear{mousavi2019stanford}) published benchmark dataset, contains a range of local seismic signals – both earthquake and non-earthquake – along with noise examples. 
The dataset includes $\sim$1.2 million waveforms, of which $\sim$200,000 are noise examples and the remaining contain seismic arrivals from $\sim$450,000 earthquakes ($\sim$M -0.5 - 8). 
The units for the waveform examples are raw counts and there are 40 metadata variables associated with this event. 

For the split, we use the same test set as defined in \cite{mousavi2020earthquake} which randomly set 10\% of the examples as testing examples, we then add a validation set by randomly sampling from the remaining samples. 
The final ratios of the training, validation, and testing split are again 60\%, 30\%, 10\% respectively. \\

The following datasets include cases where the publicly available waveform data, along with corresponding metadata was available for training ML models, but some common metadata is missing.\\

\noindent NEIC\\
\noindent The National Earthquake Information Centre (NEIC; \citeauthor{yeck2021leveraging}, \citeyear{yeck2021leveraging}) published benchmark dataset comprises $\sim$1.3 million seismic phase arrivals with global source-station paths. 
As information on the trace start-time and station is missing for this dataset, it is stored in the SeisBench format, but without this normally required information. 

For the training, development and testing split, the original publication presented randomly sampled splits,  based on event-id. 
This random splitting approach is implemented in the SeisBench conversion of this dataset, again at 60\%, 10\%, and 30\% for the training, development, and testing examples respectively.\\

There are additional datasets integrated into SeisBench which were originally used in training notable deep learning algorithms in seismology. 
Typically, the waveforms for these datasets were already pre-processed for training, including windowing and labelling, so the original station metadata for each training example is unavailable for these datasets. 
As many of the datasets also use picked waveforms from the SCEDC network, this results in potential common overlap between the following listed datasets, for both metadata parameters and waveforms.  
The only differences being potentially different metadata variables across datasets (e.g. picked phase labels, vs. first motion labels). 

The deep learning training datasets converted into SeisBench format include: 
the 'GPD' training dataset \citep{ross2018generalized} containing 4,773,750 examples of 4s waveforms, sampled at 100~Hz; 
the 'JGRFM' dataset used for training the deep learning-based first motion polarity detection routine in the \cite{ross2018p} study, containing  6~s Z-component waveform samples from 100~Hz instruments; 
the 'JGRPick' dataset used for training the deep learning-based picker presented in the same work; The 'Meier2019JGR' dataset, which contains the S. Californian component of the training examples from the \cite{meier2019reliable} work.

\section{Models}
The SeisBench \textit{model} interface is an extensible framework which encompasses the application of all types of models to seismic data. 
It is designed to be generalizable to arbitrary seismic tasks which operate on waveform data. 
A range of deep learning models from the literature are provided through SeisBench (Table 2). 
All deep learning models are integrated with the PyTorch framework \citep{paszke2017automatic}. 
Where possible, models integrated into SeisBench have the corresponding weights from the original training procedure integrated. 
We also provide weights for each of the models trained on each of the included datasets (see the companion paper to this work, \cite{munchmeyer2021benchmark}). 

\subsection{Initially integrated models}

The initial set of models integrated into SeisBench are listed below. For a more detailed description, refer to \citep{munchmeyer2021benchmark}.

\begin{table*}

\begin{center}
\footnotesize
\begin{tabular}{l|p{2cm}p{2cm}p{2cm}p{2cm}p{2cm}p{2cm}}
& BascicPhaseAE & CRED & DPP & EQT  & GPD & PhaseNet\\
\hline
\# Params & 33,687 & 293,569 & 199,731/ 546,081/ 21,181 & 376,935 & 1,741,003 & 23,305\\
Type & U-Net & CNN-RNN & CNN/RNN/RNN & CNN-RNN-Attention & CNN & U-Net\\
Training set & N. Chile & S. California & N. Chile & STEAD & S. California & N. California \\
Orig. weights & N & Y & N & Y & Y & N\\
Reference & \cite{woollam2019convolutional} & \cite{mousavi2019cred} & \cite{soto2021deepphasepick} & \cite{mousavi2020earthquake} & \cite{ross2018generalized} & \cite{zhu2019phasenet}\\
\end{tabular}
\caption{Description of the models studied. The number of parameters refers to the total number of trainable parameters. Note that these numbers might deviate slightly from the ones published by the original authors due to differences in the underlying frameworks. For DPP, information delimited by slashes indicate Detector/P-Picker/S-Picker networks. The row "Orig. weights" indicates whether original weights were published and are available in SeisBench. For PhaseNet, weights were published by the authors, but these weights could not straightforwardly be integrated into SeisBench due to technical issues.}
\label{tab:models}
\end{center}
\end{table*}

\begin{itemize}
  \item BasicPhaseAE \citep{woollam2019convolutional}, basic CNN U-Net, initially applied to regional aftershock sequence in Chile.
  \item CRED \citep{mousavi2019cred}, CNN-RNN Earthquake Detector, initially trained on 500,000 training signal and noise examples from Northern California.
  \item DPP \citep{soto2021deepphasepick}, DeepPhasePick, is a combination of a convolutional neural network for phase detection and two recurrent neural networks for onset time determination.  Like BasicPhaseAE, the networks were designed for detecting and picking local events, with an initial application on a regional seismic network in Chile.
  \item EQT \citep{mousavi2020earthquake}, EarthQuake Transformer, an Attention-based Transformer Network to both detect and pick events.
  \item GPD \citep{ross2018generalized} Generalised Phase Detection, CNN algorithm to detect seismic phases.
  \item PhaseNet \citep{zhu2019phasenet}, CNN autoencoder algorithm, adapts the U-Net segmentation framework to the 1D problem of classifying seismic phases.
\end{itemize}

\section{Training data generation pipeline}
A common task for training ML models in seismology is building data generation pipelines for training. 
First, some pre-processing is usually done; for example, traces need to be truncated to the correct length and possibly normalized, labels need to be encoded.
Furthermore, often it is beneficial to augment the data to increase the variability on training examples, 
for example by adding noise to the waveforms. 
To standardize this task, reduce the required coding amount and reduce errors in the training pipeline, SeisBench provides the \textit{generate} API (cf. Figure~\ref{fig_sb_schematic}). 

The \textit{generate} API provides individual processing blocks, e.g., window selection, label definition, or normalization, which can be combined into a data generation pipeline in a flexible way.
While many standard augmentations are already implemented, custom routines can be added easily. 
As the \textit{generate} API only relies on the abstract \textit{data} API, the same set of augmentations can be applied to any SeisBench compatible dataset with minimal changes in the code. 
In addition, since the \textit{generate} API is integrated with PyTorch, it can facilitate efficient data generation with PyTorch's built-in multi-processing.

\section{Example workflows - Using SeisBench benchmark datasets and models}

Here we highlight how the features and functionality provided through SeisBench can support users with their tasks, from practitioners just looking to use an ML model to experts wishing to conduct extensive, in-depth, comparison and benchmarking pipelines. 

\subsection{Workflow 1 - Use pre-trained models for picking new seismic streams}

\begin{figure*}
  \includegraphics[width=\textwidth]{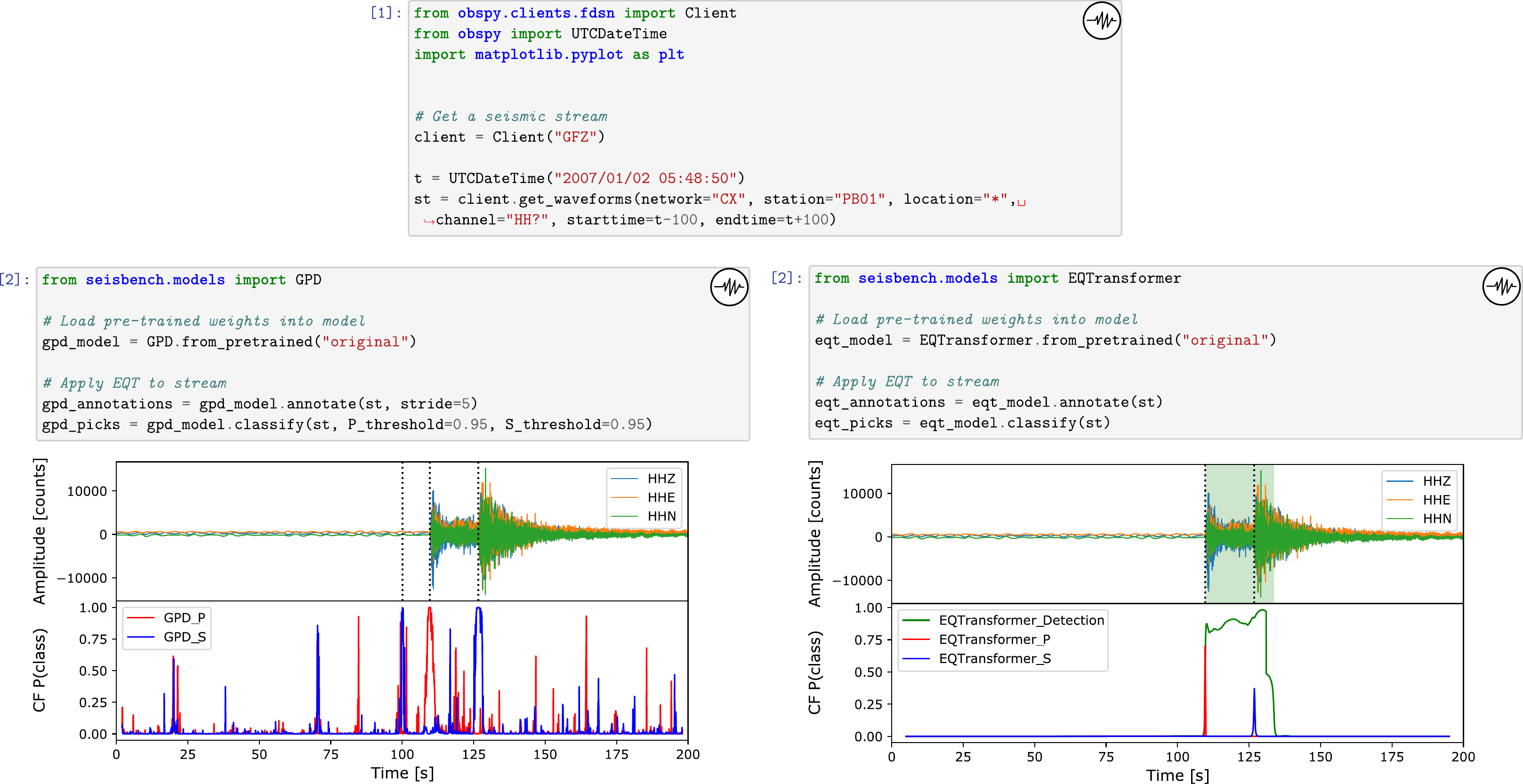}
  \caption{Example code-blocks which download a seismic waveform [1], then loads a pre-trained deep learning picking model and applies the model to predict on the seismic stream using either one of two ML architectures (GPD and EQTransformer) [2]. Resulting picks and characteristic functions from the output probabilities are displayed beneath the code blocks. Picks are represented by dotted lines, event detections for the EQT case are the shaded regions. The GPD picker makes a spurious S -pick before the onset of the event but as the original model weights have been incorporated into the pickers to pick on new, unseen data, this example may not be representative of the optimum performance of the respective model architectures, which could be achieved by training on data matched to the application case.}
\label{fig_feature_highlight}
\end{figure*}

This workflow is relevant for practitioners who seek to leverage ML techniques on seismic data, but do not necessarily have the in-depth domain knowledge to do this through ML frameworks. 
This example demonstrates how to pick seismic waveforms with two leading, pre-trained models (EQT and GPD) via the SeisBench API. 
The commands to do this are displayed in Figure \ref{fig_feature_highlight}. 
The high-level functionality allows users to apply ML models to seismic data with 
just a few commands. 
If not previously downloaded, the pre-trained model weights are downloaded and subsequently cached for repeat use. 
The annotate and classify methods of the SeisBench models integrate with stream objects from the obspy package \citep{beyreuther2010obspy}, widely used within the seismological community. 
We omit the plotting code for brevity. 
Users can easily expand upon this example workflow to conduct seismic detection and picking pipelines. 
In terms of computational performance, we test the EQTransformer implementation on a K80 GPU and annotate 24 hours of 100~Hz data from a single station in 6~s. Scaling this process results in a months worth of data being labelled in $\sim$3 minutes.

\subsection{Workflow 2 - Training models}

\subsubsection{Training a deep learning model}

\begin{figure*}
  \includegraphics[width=\textwidth]{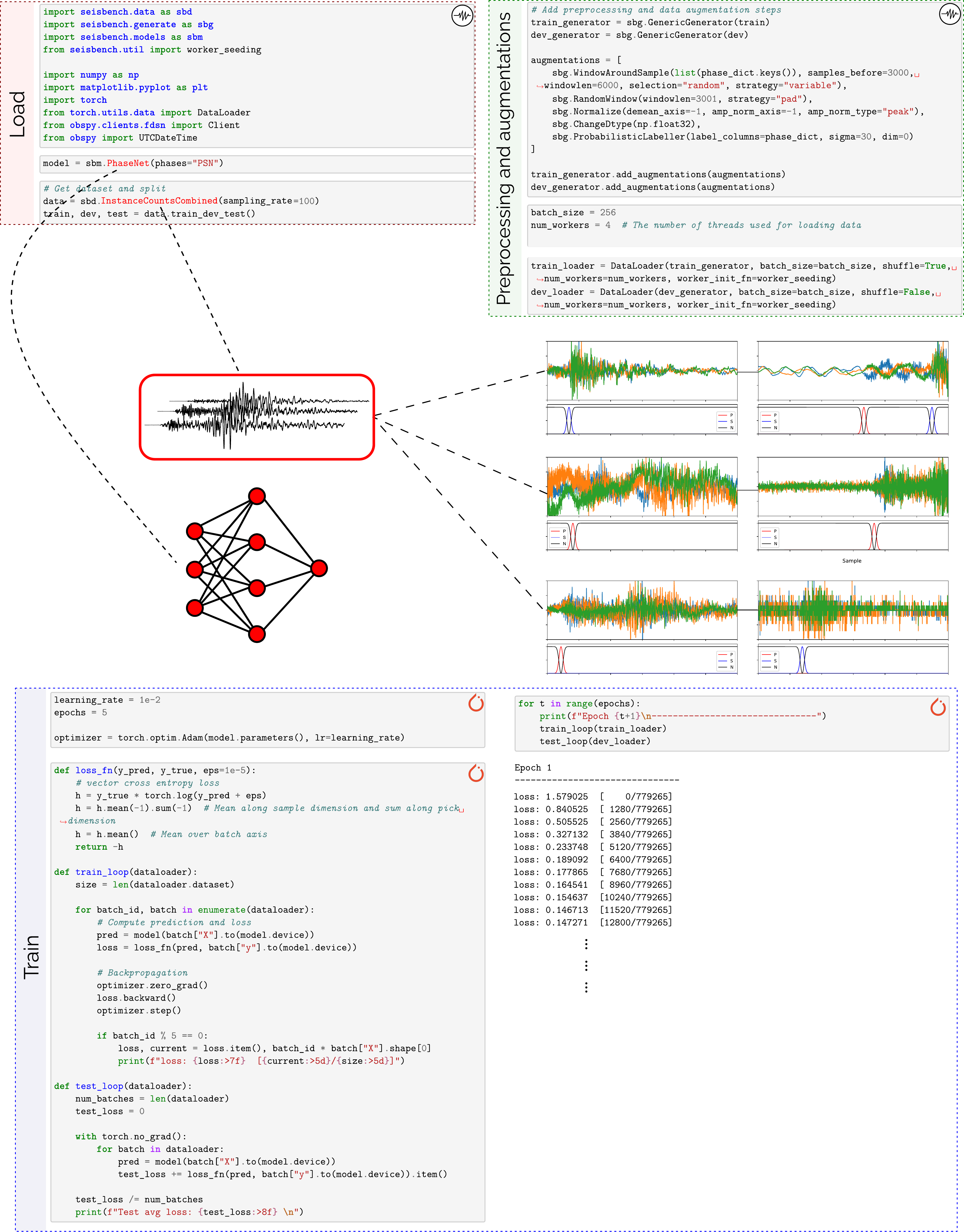}
  \caption{Example code-block with additional schematic diagrams displaying syntax required to perform full training of a deep learning model in SeisBench. PhaseNet is used for training, with the INSTANCE dataset being used as training data. Further workflow examples demonstrating the functionality provided by SeisBench can be found at \url{https://github.com/seisbench/seisbench/tree/main/examples}.}
\label{fig_feature_highlight2}
\end{figure*}

For those wishing to train a deep learning model, Figure \ref{fig_feature_highlight2} provides a run through of how this workflow can be built in SeisBench. 
This workflow highlights how the \textit{data}, \textit{generate}, and \textit{model} modules combine to help users perform all the typical tasks required in such a pipeline.  
Any loading of the required models and data is performed initially. 
In this example, we train PhaseNet on the INSTANCE dataset. 
Once the dataset and model are loaded, the \textit{generate} module can be used to perform typical pre-processing and data augmentation steps on the waveforms. 
The generator object accepts a suite of augmentations which will be applied to each batch during training. 
In this example, we randomly window the waveforms, normalise the amplitudes, change the datatype to 32bit floats, finally creating a probabilistic vector representation of P-picks, S-picks, and noise examples in the waveform. 
These steps can be achieved in 10 lines of code through SeisBench (see the \emph{Preprocessing and augmentations} code block, Figure \ref{fig_feature_highlight2}).
The waveforms following processing are displayed in Figure \ref{fig_feature_highlight2}. 
The augmented waveform data then form a training sample for PhaseNet. 
We also show the standard PyTorch syntax to iterate through a DataLoader object and train the model as the last step (see the \emph{Train} code block, Figure \ref{fig_feature_highlight2}).

\subsubsection{Transfer learning}
Rather than train a new model from scratch, transfer learning forms another common workflow users may require. 
The modular nature of the API means that to switch any dataset or model for another, all that is required is to change data or model imported (indicated by the dashed lines in Figure \ref{fig_feature_highlight2}). 
To load a pre-trained model, all users have to do is call the \textit{from\textunderscore pretrained} method. 
The syntax to perform this step is also displayed in workflow 1. 
Datasets can also be swapped easily. For the purposes of this example, any dataset containing P-, and S-picks could be loaded in place of the INSTANCE dataset in workflow 2, and the training would then be performed on this alternative dataset, using the PhaseNet model initially trained on regional seismic waveforms in California as initialization for the training.

\subsection{Workflow 3 - Benchmark differing models across differing datasets}
Beyond training for a single model or dataset, SeisBench allows for comparison pipelines to be easily constructed.
Having an objective measure of the performance of newly proposed algorithms against current state-of-the-art routines is fundamental to progress in any field, and standard procedure in traditional ML domains such as image recognition. 
As ML is a recent adoption within seismology, it could be argued that this step has not yet been carried out extensively. 
A detailed benchmarking study of various published ML picking models was carried out by us with the SeisBench framework and is presented with the companion paper to this work \citep{munchmeyer2021benchmark}. The code used for this benchmark study is made available and can serve as a template for future benchmarking studies\footnote{Available at \url{https://github.com/seisbench/pick-benchmark}}.

\section{Extensibility}
The SeisBench API is published with an open-source license (GPLv3). 
The software is designed to be extensible, and we encourage the seismological community to contribute. 
If users wish to integrate their own benchmark datasets or models to the package for public download, we ask that they get in touch with the project through GitHub (\url{https://www.github.com/seisbench/seisbench}); where further information on the contribution guidelines can be found.
In particular, we encourage inclusion of already published models and datasets. 
The code-base has extensive test coverage to reduce the risk of coding errors.

With the picking and detection problems having been widely explored in recent years with ML approaches,  more complex problems are now being tackled with these techniques. 
We envisage that the models incorporated into SeisBench will expand to include such tasks. 
For example, hypocentre determination, source parameter estimation, etc., can all be constructed with SeisBench. 
All that is required is that the labels for a supervised learning task are present in the metadata. 
Once the state-of-the-art ML models for a given task are available in SeisBench - as shown with the picking example above - the major advantages of integrating new models within
this framework become apparent. 
The initial processing routine set up for a  model can be directly used to compare against existing  state-of-the-art models. 
This ease of testing will hopefully promote further innovation of ML in seismology. 

\section{Conclusions}
 We have developed SeisBench as an open-source Python package, built to aid users in their application of ML techniques to seismic data. 
 It minimizes common barriers to development for both practitioners looking to apply ML methods to seismic tasks, and experts who wish to benchmark and train leading algorithms. 
 The software provides access to recently published benchmark datasets for machine learning in seismology, downloadable and accessible through a common interface. 
 SeisBench extends this concept to provide a common access point to ML models, 
 with state-of-the-art models and corresponding weights for seismic tasks directly integrated. 
 We provide access to a range of picking models from the literature in the first iteration of the software but the framework is applicable for many seismological tasks based on waveform analysis such as location and magnitude estimation. 
 By tackling some of the common bottlenecks encountered when developing ML algorithms, we hope that SeisBench will help practitioners iterate and deploy their models, advancing the development of the next generation of ML techniques within seismology.

\section*{Acknowledgements}

We thank the Impuls- und Vernetzungsfonds of the HGF to support the REPORT-DL project under the grant agreement ZT-I-PF-5-53. We use PyTorch \citep{paszke2017automatic} for integrating deep learning models into the package. JM acknowledges the support of the Helmholtz Einstein International Berlin Research School in Data Science (HEIBRiDS).
This work was also partially supported by the project INGV Pianeta Dinamico 2021 Tema 8 SOME (CUP D53J1900017001) funded by Italian Ministry of University and Research ``Fondo finalizzato al rilancio degli investimenti delle amministrazioni centrali dello Stato e allo sviluppo del Paese, legge 145/2018''.

We would also like to thank the authors of the original benchmarking and model papers, who made publicly available either benchmark datasets, or original model weights. 
Open-source development is an important component of advancing research and SeisBench only works as a tool with these models and datasets openly accessible.

\normalsize
\bibliography{bibliography}

\clearpage
\appendix

\setcounter{table}{0}
\renewcommand{\thetable}{A\arabic{table}}
\onecolumn

\begin{longtable}{p{5.5cm}|p{11cm}} 
Parameter name & Description\\
\hline
trace\textunderscore name & A unique identifier for the trace.\\
trace\textunderscore start\textunderscore time & The start time for the trace - if possible following ISO 8601:2004.\\
trace\textunderscore sampling\textunderscore rate\textunderscore hz & Sampling rate of the trace. If sampling rate is constant across all traces in the data set, it can also be specified in the data\textunderscore format group in the hdf5 data file.\\
trace\textunderscore npts & The number of samples in the trace.\\
trace\textunderscore channel & The channel from which the data was obtained without the component identifier, e.g., HH, HN, BH.\\
trace\textunderscore category & A category to assign to the trace, e.g. earthquake, noise, mine blast.\\
trace\textunderscore snr\textunderscore db & The signal-to-noise ratio of trace in decibels.\\
trace\textunderscore p\textunderscore arrival\textunderscore sample & Sample in trace at which P-phase arrives\\
trace\textunderscore p\textunderscore uncertainty\textunderscore s & Uncertainty of P-phase pick in seconds.\\
trace\textunderscore p\textunderscore weight & The weighting factor assigned to the P-phase pick.\\
trace\textunderscore p\textunderscore status & The status of the P-phase pick, e.g. manual/automatic.\\
trace\textunderscore s\textunderscore arrival\textunderscore sample & Sample in trace at which S-phase arrives.\\
trace\textunderscore s\textunderscore uncertainty\textunderscore s & Uncertainty of S-phase pick in seconds.\\
trace\textunderscore s\textunderscore weight & The weighting factor assigned to the S-phase pick.\\
trace\textunderscore s\textunderscore status & The status of the S-phase pick, e.g. manual/automatic.\\
trace\textunderscore completeness & The fraction of samples in the trace, which were not filled with placeholder values (between 0 and 1). Placeholder values occur for example in case of recording gaps or missing component traces.\\
\hline
source\textunderscore id & A unique identifier for the source trace.\\
source\textunderscore origin\textunderscore time & Origin time of the source - if possible following ISO 8601:2004.\\
source\textunderscore origin\textunderscore uncertainty\textunderscore sec & Uncertainty of source origin time in seconds.\\
source\textunderscore latitude\textunderscore deg & Source latitude coordinate in degrees.\\
source\textunderscore latitude\textunderscore uncertainty\textunderscore deg & Uncertainty of source latitude coordinate in degrees.\\
source\textunderscore longitude\textunderscore deg & Source longitude coordinate in degrees.\\
source\textunderscore longitude\textunderscore uncertainty\textunderscore deg & Uncertainty of source longitude coordinate in degrees.\\
source\textunderscore depth\textunderscore km & Source depth in kilometers.\\
source\textunderscore depth\textunderscore uncertainty\textunderscore km & Uncertainty of source depth coordinate in degrees.\\
source\textunderscore error\textunderscore sec & The error association with the source location in seconds.\\
source\textunderscore gap\textunderscore deg & Azimuthal gap from the source determination in degrees.\\
source\textunderscore magnitude & Magnitude value assigned to source.\\
source\textunderscore magnitude\textunderscore type & The type of magnitude calculation used when assigning magnitude to source.\\
\hline
station\textunderscore network\textunderscore code & Instrument network code.\\
station\textunderscore code & Instrument station code.\\
station\textunderscore location\textunderscore code & Instrument location code.\\
station\textunderscore latitude\textunderscore deg & Instrument latitude in degrees.\\
station\textunderscore longitude\textunderscore deg & Instrument longitude in degrees.\\
station\textunderscore elevation\textunderscore m & Instrument elevation in m.\\
\hline
path\textunderscore back\textunderscore azimuth\textunderscore deg & The backazimuth of phase path from source to receiver in degrees.\\
path\textunderscore ep\textunderscore distance\textunderscore km & The epicentral distance of source receiver path in kilometers.\\
path\textunderscore hyp\textunderscore distance\textunderscore km & The hypocentral distance of source receiver path in kilometers.\\
path\textunderscore p\textunderscore travel\textunderscore s & Travel-time for P-phase in seconds.\\
path\textunderscore p\textunderscore residual\textunderscore s & Residual of P-phase against some prediction in seconds.\\
path\textunderscore s\textunderscore travel\textunderscore s & Travel-time for P-phase in seconds.\\
path\textunderscore s\textunderscore residual\textunderscore s & Residual of S-phase against some prediction in seconds.\\
\caption{Example of the parameter naming schema for SeisBench, where metadata parameters follow the naming format guidelines ‘CATEGORY\textunderscore PARAMETER\textunderscore UNIT’. The table displays a subset of some of the more common naming parameters. When extending or including new datasets, this can be extended for individual use cases to include any new metadata parameter, providing it adheres to the naming schema.}
\label{param_description_table}
\end{longtable}

\end{document}